\documentclass[aps,epsfig,amssymb,twocolumn]{revtex4}
\begin{document}
\title{No-deleting and no-cloning principles as  consequences of conservation of
quantum information}
\author{Micha{\l} Horodecki, Ryszard Horodecki, Aditi Sen(De) and Ujjwal Sen}
\affiliation{Institute of Theoretical Physics and Astrophysics, University of Gda\'nsk, 
         80-952 Gda\'nsk, Poland}

\begin{abstract}

We show that the no-deleting and no-cloning principles
are implications of information  conservation principle.
This is unlike in classical physics, where cloning and  deleting
are possible, independently of information conservation.
Connections with the second law of thermodynamics are also discusssed.

\end{abstract}

\maketitle

\def\com#1{{\tt [\hskip.5cm #1 \hskip.5cm ]}}
\newcommand{\tr}{{\rm tr}}

\section{Introduction}

According to the generic information paradigm \cite{H91, HH_basic} all physical processes
in nature should be formulated in terms of processing of information, the latter being
a fundamental quantity that can be defined independently of probability itself \cite{Kolmogorow,
S, C, IngardenU, IngardenK}.
This program got an excellent ``tool" - the quantum information paradigm (see
e.g. \cite{Jozsa_recent, HHH_recent}).
Then it seems  resonable to reformulate the quantum formalism in terms of
quantum
information. Such a task naturally requires some axiomatic approach to the
latter.

In this spirit, the conservation of quantum information was proposed
\cite{HH_basic, HHH_balance} as a basic principle in processing of quantum information. It can be
formulated as follows: 

``In a closed system, one cannot change entanglement by
local operations.''

Note that this formulation is ``quantum", as it does not
say about probabilities but merely about entanglement - a purely quantum
property of compound systems.
   In this context it is natural to ask: \emph{What are the fundamental consequences
of the conservation  of information principle?} In this paper we show that
both  no-deleting and no-cloning principles, being   consequences
\cite{Braunstein_Pati_deleting,WoottersZurek,Dieks}
of the quantum
formalism,
 are implications of conservation of quantum information. It is also
suggested that the converse implication holds. We also discuss the  basic
relations between the no-cloning and no-deleting principles and second law
of thermodynamics, the latter being treated as
part of the information conservation principle.

\section{The no-deleting \emph{principle}}
The no-deleting principle states that in a closed system, one cannot detroy 
quantum information. In closed systems, quantum information can only be moved from one 
place (subspace) to another. One formulation of this principle is the following:

Given two qubits (two-dimensional
quantum systems, say an electron or a photon)
 in arbitrary but equal states, one cannot take one of them to a 
fixed state, while keeping the other in the original state. More formally, one cannot 
have the evolution
\[
\left|\psi\right\rangle \left|\psi\right\rangle \rightarrow 
\left|\psi\right\rangle \left|0\right\rangle,
\]
where 
\(\left|\psi\right\rangle\) is an arbitrary state of a qubit and 
\(\left|0\right\rangle\) is a fixed state. 

A sharper formulation is that it is not possible to delete 
a copy against another when given from among two nonorthogonal states. More formally,
the evolution
\begin{equation}
\label{nonorthogonal_deleting}
\begin{array}{lcl}
\left|\psi_1\right\rangle \left|\psi_1\right\rangle \rightarrow 
\left|\psi_1\right\rangle \left|0\right\rangle,\\
\left|\psi_2\right\rangle \left|\psi_2\right\rangle \rightarrow 
\left|\psi_2\right\rangle \left|0\right\rangle,
\end{array}
\end{equation}
is not possible,
where 
\(\left\langle \psi_1 | \psi_2 \right\rangle \ne  0\).

The above statements of the no-deleting principle can be proven as theorems, by assuming
unitary evolution. This was done in \cite{Braunstein_Pati_deleting}, and was 
called the no-deleting \emph{theorem}.

Note the importance of deleting one copy \emph{against another copy}. 
In a closed system, even orthogonal states cannot be deleted \emph{alone}, 
if an unitary evolution 
is assumed. One cannot take the orthogonal 
states \(\left|0\right\rangle\) and \(\left|1\right\rangle\) to 
a standard state \(\left|0\right\rangle\), even with ancillas. If one assumes
that 
\[
\begin{array}{lcl}
\left|0\right\rangle \left|A\right\rangle \rightarrow 
                        \left|0\right\rangle \left|A_0\right\rangle,\\
\left|1\right\rangle \left|A\right\rangle \rightarrow 
                          \left|0\right\rangle \left|A_1\right\rangle
\end{array}
\]
is true under an unitary evolution, the only option is to have 
orthogonal \(\left|A_0\right\rangle\)
and \(\left|A_1\right\rangle\). That is the information is merely transfered from the system
to the ancilla, and not deleted. Therefore even classical information cannot be deleted
\emph{alone} in unitary evolution. In contrast, classical information can of course 
be deleted \emph{against a copy}. Therefore 
the transformations 
\[
\begin{array}{lcl}
\left|0\right\rangle \left|0\right\rangle \rightarrow 
                        \left|0\right\rangle \left|0\right\rangle,\\
\left|1\right\rangle \left|1\right\rangle \rightarrow 
                          \left|1\right\rangle \left|0\right\rangle
\end{array}
\]
is possible by a single unitary evolution, for example by the so-called CNOT gate.

More important is to realize that the no-deleting principle is true only in 
\emph{closed} systems 
(see \cite{Josza_strong_cloning} in this regard). One can of course delete 
a state of a system, even alone, by simply 
throwing it out of the system. A more ``sophisticated'' way of saying it is the following:
Given a qubit in a state \(\left|\psi\right\rangle\), an observer (call him
``Demon'') performs a measurement in the 
computational basis \(\{\left|0\right\rangle\), \(\left|1\right\rangle\}\). 
If the result is  \(\left|0\right\rangle\), Demon keeps the output as 
\(\left|0\right\rangle\). However if the result is \(\left|1\right\rangle\),  Demon
flips (by applying, for example, 
\(\sigma_x=\left(
\begin{array}{lcl}
 0 && 1\\
 1 && 0
\end{array}
\right)\), written in the computational basis) the output \(\left|1\right\rangle\) to 
\(\left|0\right\rangle\). Thus Demon is able to delete an arbitrary 
qubit to a standard state. And actually he is able to do it for a single 
copy alone, which we have seen to be not possible even for classical information if the 
system is closed. But the system here is open. The information about 
the result of measurement is now in Demon's ``head''. Formally, 
the superoperator (trace-preserving completely positive map) that effects 
the above transformation from \(\left|\psi\right\rangle\) to 
\(\left|0\right\rangle\) is the following:
\[
\begin{array}{lcl}
\rho \rightarrow \Lambda  (\rho) &=& I \left|0\right\rangle \left\langle 0 \right|
                                     \rho \left|0\right\rangle \left\langle 0 \right| I
+  \sigma_x \left|1\right\rangle \left\langle 1 \right|
                            \rho \left|1\right\rangle \left\langle 1 \right| \sigma_x \\
&=& \left|0\right\rangle \left\langle 0 \right|
                                     \rho \left|0\right\rangle \left\langle 0 \right| 
+  \left|0\right\rangle \left\langle 1 \right|
                            \rho \left|1\right\rangle \left\langle 0 \right|. 
\end{array}
\]
(Here \(I\) denotes the identity operator on the qubit Hilbert space.) Every superoperator
can be seen as an unitary evolution in an extended Hilbert space. Specifically,
any superoperator evolution \(\rho \rightarrow \sum_i A_i \rho A_i^\dagger\), with 
\(\sum_i A_i^\dagger A_i\) being the identity operator on the Hilbert space on which \(\rho\) 
is defined, can be seen as the unitary evolution \(\left|
\phi \right\rangle \left| 0\right\rangle_E \rightarrow \sum_i 
A_i \left| \phi\right\rangle \left|i\right\rangle_E \), for orthogonal states 
\(\left|i\right\rangle\) of 
the ``environment'' E. (Here \(\left|\phi\right\rangle\) is an arbitrary 
 vector from the Hilbert space
on which \(\rho\) is defined.) In our case,  the superoperator 
\(\Lambda\) can be seen as the following  unitary evolution:
\[
\left|\psi\right\rangle \left|0\right\rangle_E \rightarrow
\left|0\right\rangle \left\langle 0 | \psi \right\rangle \left|0\right\rangle_E
+ \left|0\right\rangle \left\langle 1 | \psi \right\rangle \left|1\right\rangle_E.   
\] 
However the right hand side is simply 
\[
\left|0\right\rangle (\left\langle 0 | \psi \right\rangle \left|0\right\rangle_E
+ \left\langle 1 | \psi \right\rangle \left|1\right\rangle_E) 
= \left|0\right\rangle \left|\psi\right\rangle_E.
\]
So the system is simply thrown out into the environment to effect the deletion. That is 
in doing the deletion, Demon simply transfered the \emph{whole} information (about 
the state to be deleted) into his 
``head''.

Thus we see that the very notion of the  no-deleting principle, naturally leads us to 
consider it in a scenario of a closed system. This (and more particularly, 
the action of Demon above) reminds us of the second law of thermodynamics. 
The whole point in exorcising the Maxwell's demon
(and save the second law of thermodynamics) 
was to include the measuring device into the system, that is to consider everything 
in a closed system (see \cite{Plenio_Vitelli} for a recent review).

\section{The no-cloning \emph{principle}}

The no-cloning principle states that quantum information cannot be ``leaked out'' while 
keeping the original information content intact. And this is true even in open systems. 

More formally, given a qubit in an unknown  state 
\(\left|\psi\right\rangle\), one cannot
obtain
\[
\left|\psi\right\rangle \left|0\right\rangle \rightarrow 
\left|\psi\right\rangle \left|\psi\right\rangle,
\]
even for open systems. Considering the environment inside the dynamics, this 
states that one cannot obtain 
\[
\left|\psi\right\rangle \left|0\right\rangle \left|0\right\rangle_E \rightarrow 
\left|\psi\right\rangle \left|\psi\right\rangle \left|e_{\psi}\right\rangle_E.
\]

Again a sharper formulation is possible, viz. that one cannot have 
\begin{equation}
\label{nonorthogonal_cloning}
\begin{array}{lcl}
\left|\psi_1\right\rangle \left|0\right\rangle \rightarrow 
\left|\psi_1\right\rangle \left|\psi_1\right\rangle  ,\\
\left|\psi_2\right\rangle \left|0\right\rangle  \rightarrow 
\left|\psi_2\right\rangle \left|\psi_2\right\rangle ,
\end{array}
\end{equation}
for nonorthogonal \(\left|\psi_1\right\rangle \) and \(\left|\psi_2\right\rangle \).
Including the environment, this implies that 	
\begin{equation}
\label{nonorthogonal_cloning_environ}
\begin{array}{lcl}
\left|\psi_1\right\rangle \left|0\right\rangle \left|0\right\rangle_E \rightarrow 
\left|\psi_1\right\rangle \left|\psi_1\right\rangle \left|e_{\psi_1}\right\rangle_E ,\\
\left|\psi_2\right\rangle \left|0\right\rangle \left|0\right\rangle_E \rightarrow 
\left|\psi_2\right\rangle \left|\psi_2\right\rangle \left|e_{\psi_2}\right\rangle_E,
\end{array}
\end{equation}
is not possible by a single evolution.

Again assuming an unitary dynamics, these statements can be proven. This was done in 
Refs. \cite{WoottersZurek,Dieks}, and is called the no-cloning \emph{theorem}.

In the following we raise this theorem to a principle and look for its connections with 
thermodynamics.

\section{The no-deleting and no-cloning principles and the second law}

In classical physics, both cloning and deleting are possible. Independently
of these,
the second law of thermodynamics holds,
 which says that entropy cannot be decreased in a closed
system.
A different formulation of second law would be that  the probabilities do
not change,
under Hamiltonian evolution in phase space.  The latter could be called
conservation of classical
information, and it includes both no-increasing of entropy as well as {\it
no-decreasing}
of entropy.

In quantum mechanics, one \emph{cannot} either delete or clone. The second
law holds too:
entropy cannot be decreased in a closed system.

Again the second law can be stated in a stronger way: that \emph{eigenvalues} of density
matrix do not change under quantum evolution. And this is actually the information
conservation principle.  


Since the second law holds for classical systems independently of deleting and
cloning, one could expect
that no-cloning and no-deleting principles will not be connected with this law, and with
law of conservation of information.

However, as we have stressed above, the  no-deleting principle has connections with the Maxwell's demon approach to
thermodynamics.
Namely, it needs a closed system: One can always delete, by throwing away the
copy. However the
copy will reside somewhere in the world.
This is similar to the second law, which says that you cannot decrease
entropy in a closed system.
In open systems,  one can do that: By bringing in  a pure state, and removing the initial
state.

Below we will argue that both no-cloning and no-deleting principles
 are consequences of
conservation of information.



\subsection{The second law contains the no-deleting principle}

In the following, we raise 
 the statement of the no-deleting theorem to the level of a \emph{principle},
and do not assume an unitary dynamics. We will 
then show that the no-deleting principle is contained in the second law of thermodynamics.

Suppose that 
deletion is possible. Then one can effect the following evolution
(call it the ``deleting evolution'') in a closed system:
\[
\left|\psi\right\rangle \left|\psi\right\rangle \rightarrow 
\left|\psi\right\rangle \left|0\right\rangle,
\]
for an arbitrary qubit \(\left|\psi\right\rangle\) and a standard state
\(\left|0\right\rangle\). For arbitrary \(\left|\psi\right\rangle\), the states
\(\left|\psi\right\rangle \left|\psi\right\rangle\) span a three-dimensional subspace
(the so-called symmetric subspace) of the four-dimensional Hilbert space of two qubits. 
However the states \(\left|\psi\right\rangle \left|0\right\rangle\) span only a 
two-dimensional subspace, as \(\left|0\right\rangle \) is a fixed state. Thus the 
deleting evolution takes a system with von Neumann entropy \(\log_2 3\) (the identity 
operator on the symmetric subspace of two qubits has von Neumann entropy \(\log_2 3\)
\cite{von_Neumann})
to one with von Neumann entropy \(\log_2 2\). Since the system is closed, 
this decrease of entropy is therefore a violation of the second law of thermodymanics. 
In other words, the second law of thermodynamics implies the no-deletion principle.

One knows that the statement of the second law of thermodynamics can be that 
``entropy does not \emph{change} in a closed system''. In this paper, 
unless stated otherwise, we will use only 
the version stating that ``entropy does not \emph{decrease} in a closed system''. 
This is the version that we have used above in showing that the second law implies the 
no-deleting principle.

The sharper form of the deleting, viz.
\[
\begin{array}{lcl}
\left|\psi_1\right\rangle \left|\psi_1\right\rangle \rightarrow 
\left|\psi_1\right\rangle \left|0\right\rangle,\\
\left|\psi_2\right\rangle \left|\psi_2\right\rangle \rightarrow 
\left|\psi_2\right\rangle \left|0\right\rangle,
\end{array}
\]
for \(\left\langle \psi_1 | \psi_2 \right\rangle \ne  0\), also violates the second law.
This is because \(\left|\psi_1\right\rangle \left|\psi_1\right\rangle \)
and \(\left|\psi_2\right\rangle \left|\psi_2\right\rangle \) are 
more farther apart than \(\left|\psi_1\right\rangle \left|0\right\rangle \)
and \(\left|\psi_2\right\rangle \left|0\right\rangle\) and consequently 
the average input state 
\(\frac{1}{2}(\left|\psi_1\right\rangle \left|\psi_1\right\rangle 
\left\langle \psi_1 \right| \left\langle \psi_1 \right|
+ \left|\psi_2\right\rangle \left|\psi_2\right\rangle 
\left\langle \psi_1 \right| \left\langle \psi_1 \right|)\) 
has less von Neumann entropy than the average output state
\(\frac{1}{2}(\left|\psi_1\right\rangle \left|0\right\rangle 
\left\langle \psi_1 \right| \left\langle 0 \right|
+ \left|\psi_2 \right\rangle \left|0\right\rangle 
\left\langle \psi_2 \right| \left\langle 0 \right|)\).

One may also see deletion in the following way \cite{unitary_deleting}:
\[
\begin{array}{lcl}
\left|\psi_1\right\rangle \left|\psi_1\right\rangle \rightarrow 
                   \left|\psi_1\right\rangle \left|a_{\psi_1}\right\rangle,\\
\left|\psi_2\right\rangle \left|\psi_2\right\rangle \rightarrow 
                     \left|\psi_2\right\rangle \left|a_{\psi_2}\right\rangle,
\end{array}
\]
where \(\left\langle \psi_1 | \psi_2 \right\rangle \ne  0\), and 
\(\left|a_{\psi_1}\right\rangle\) and \(\left|a_{\psi_2}\right\rangle\)
are more nearer than \(\left|\psi_1\right\rangle\) and \(\left|\psi_2\right\rangle\).
That is, \(|\left\langle \psi_1 | \psi_2 \right\rangle| < 
|\left\langle a_{\psi_1} | a_{\psi_2} \right\rangle|\). Again we see that 
the average input state 
\(\frac{1}{2}(\left|\psi_1\right\rangle \left|\psi_1\right\rangle 
\left\langle \psi_1 \right| \left\langle \psi_1 \right|
+ \left|\psi_2\right\rangle \left|\psi_2\right\rangle 
\left\langle \psi_1 \right| \left\langle \psi_1 \right|)\) 
has less von Neumann entropy than the average output state
\(\frac{1}{2}(\left|\psi_1\right\rangle \left|a_{\psi_1}\right\rangle 
\left\langle \psi_1 \right| \left\langle a_{\psi_1} \right|
+ \left|\psi_2 \right\rangle \left| a_{\psi_2}  \right\rangle 
\left\langle \psi_2 \right| \left\langle a_{\psi_2} \right|)\).

Therefore whatever is the form of deletion, it violates the second law of thermodynamics.



\subsection{The second law for open systems contains the no-cloning principle}

Can the no-cloning principle be also implied by the second law? To answer this question, we 
have first of all to formulate the second law of thermodynamics for \emph{open}
systems (see for example, \cite{Alickibook}). For open systems, the second law can be 
stated as follows:

The relative entropy distance between two states \(\rho\) and \(\sigma\)
does not increase.

The relative entropy distance of  \(\rho\) from \(\sigma\)
is denoted as \(S(\rho|\sigma)\) as defined as \(\tr (\rho \log_2 \rho
- \rho \log_2 \sigma)\) (which is in fact the Lindblad-Uhlmann monotonicity, see 
e.g. \cite{Ohya-Petz}).

Suppose now that the transformation  in eq. (\ref{nonorthogonal_cloning_environ}) is possible.
Then the average relative entropy distance of the input states from the average 
input state \(\rho_{in} = \frac{1}{2}(\left|\psi_1\right\rangle \left|0\right\rangle \left|0\right\rangle 
\left\langle \psi_1 \right| \left\langle 0 \right| \left\langle 0 \right|
+ \left|\psi_2 \right\rangle \left|0\right\rangle \left|0\right\rangle 
\left\langle \psi_2 \right| \left\langle 0 \right| \left\langle 0 \right|)\) is 
\[
\begin{array}{lcl}
\frac{1}{2}(S(\left|\psi_1\right\rangle \left|0\right\rangle \left|0\right\rangle
\left\langle \psi_1 \right| \left\langle 0 \right| \left\langle 0 \right||\rho_{in})\\
+ 
S(\left|\psi_2\right\rangle \left|0\right\rangle \left|0\right\rangle 
\left\langle \psi_2 \right| \left\langle 0 \right|\left\langle 0 \right||\rho_{in})) \\
= S(\rho_{in}).
\end{array}
\]
However this quantity (which is actually the Holevo quantity for the input 
ensemble \cite{Holevo}) is less than 
the average relative entropy distance of the output states from the average 
output state \(\rho_{out}=\frac{1}{2}(\left|\psi_1\right\rangle \left|\psi_1\right\rangle \left|e_{\psi_1}\right\rangle 
\left\langle \psi_1 \right| \left\langle \psi_1 \right| \left\langle e_{\psi_1} \right|
+ \left|\psi_2\right\rangle \left|\psi_2\right\rangle \left|e_{\psi_2}\right\rangle 
\left\langle \psi_1 \right| \left\langle \psi_1 \right| \left\langle e_{\psi_2} \right|)\) is 
\[
\begin{array}{lcl}
\frac{1}{2}(S(\left|\psi_1\right\rangle \left|\psi_1\right\rangle \left|e_{\psi_1}\right\rangle 
\left\langle \psi_1 \right| \left\langle \psi_1 \right| \left\langle e_{\psi_1} \right||\rho_{out}) \\
+ 
S(\left|\psi_2\right\rangle \left|\psi_2\right\rangle \left|e_{\psi_2}\right\rangle
\left\langle \psi_2 \right| \left\langle \psi_2 \right|\left\langle e_{\psi_2} \right||\rho_{out})) \\
= S(\rho_{out}).
\end{array}
\]

Therefore the second law of thermodynamics for open systems contains the no-cloning 
principle.

Again one can consider cloning in the following way:
\[
\begin{array}{lcl}
\left|\psi_1\right\rangle  \left|0\right\rangle_E \rightarrow 
\left|\psi_1\right\rangle  \left|e_{\psi_1}\right\rangle_E ,\\
\left|\psi_2\right\rangle  \left|0\right\rangle_E \rightarrow 
\left|\psi_2\right\rangle  \left|e_{\psi_2}\right\rangle_E,
\end{array}
\]
where \(\left\langle \psi_1 | \psi_2 \right\rangle \ne 0\) and the environmental states 
\(\left|e_{\psi_1}\right\rangle_E \) and \(\left|e_{\psi_2}\right\rangle_E \) are different. That is some information
has ``leaked out'' into the environment, while keeping the original copies undisturbed \cite{unitary_cloning,BBM}.
A similar argument as above shows that even this weaker form of cloning violates the second law.  
Note that in this case, second law is not non-decrease of entropy but non-increase!


\section{The no-deleting and no-cloning principles and conservation of
information}

In the following, we will show that in any physical system, conservation of information 
(in the sense of conservation of entanglement under local operations) implies
both no-deleting and no-cloning principles.

Let us first specifically state what we mean by a physical system:
\begin{enumerate}
\item The dynamics of the system is linear on the level of pure states (this can of course be
easily generalised to mixed states),
\item The system is always in a pure state,
\item Addition of ancillas is allowed.
\end{enumerate}

We do not assume unitary dynamics. And we will explicitly state when 
we allow tracing out as a valid operation. 

We will show that in such physical systems,
\begin{enumerate}
\item Conservation of information (actually, no-\emph{decrease} of entanglement under local 
operations) implies the no-deleting principle
\item Conservation of information (actually, no-\emph{increase} of entanglement under local 
operations) implies the no-cloning principle
\end{enumerate}
In the second implication, we allow tracing out as a valid operation.

Note that it is natural that we do not allow tracing out as a valid operation in considerations
of the no-deleting principle. When we are trying to show that reducing
 information from the system is not possible, we 
cannot afford to be ``careless'' and allow throwing out parts of the system. On the 
other hand, in considerations with no-cloning, when we are trying to show that 
production of information in a system is not possible, we must allow tracing out. After all,
throwing out part of a system cannot \emph{produce} information.

\subsection{No-decrease of entanglement in a closed 
system implies the no-deleting principle}

We show here that the no-deleting principle is  implied by the principle 
 that entanglement cannot decrease 
in a closed bipartite system.

By entanglement for a bipartite pure state \(\left|\alpha_{AB} \right\rangle\), we mean
the von Neumann entropy of the local density matrix 
\(\tr_{A or B} \left|\alpha_{AB} \right\rangle \left\langle \alpha_{AB} \right|\)
of \(\left|\alpha_{AB} \right\rangle\) \cite{entanglement}.

Again we prove by contradiction.
Suppose therefore  that an observer, Alice, prepares the state 
\[\left|\Psi_{AB}\right\rangle = 
\frac{1}{\sqrt{2}}(\left|0\right\rangle \left|\psi_1\right\rangle \left|\psi_1\right\rangle
+ \left|1\right\rangle \left|\psi_2\right\rangle \left|\psi_2\right\rangle)
\]
and sends the second and third systems to a distant party, Bob,  while keeping the first 
system to herself.

If deleting is possible, that is if there is a machine (available to Bob) 
that effects the transformation 
in eq. (\ref{nonorthogonal_deleting}), then Bob can locally change the state 
\(\left|\Psi_{AB}\right\rangle\) into the state 
\[\left|\Psi{'}_{AB}\right\rangle = 
\frac{1}{\sqrt{2}}(\left|0\right\rangle \left|\psi_1\right\rangle \left|0\right\rangle
+ \left|1\right\rangle \left|\psi_2\right\rangle \left|0\right\rangle).
\]
Clearly, \(\left|\Psi{'}_{AB}\right\rangle\) possesses less entanglement than 
\(\left|\Psi_{AB}\right\rangle\), for nonorthogonal 
\(\left|\psi_1\right\rangle\) and \(\left|\psi_2\right\rangle\).

\subsection{No-increase of entanglement implies the no-cloning principle}

We now show that the no-cloning principle is implied by the principle that 
 entanglement cannot increase 
in bipartite systems (see \cite{HH_basic} in this regard).

Suppose again that 
Alice, prepares the state 
\[\left|\Phi_{AB}\right\rangle = 
\frac{1}{\sqrt{2}}(\left|0\right\rangle \left|\psi_1\right\rangle \left|0\right\rangle
                                                            \left|0\right\rangle
+ \left|1\right\rangle \left|\psi_2\right\rangle \left|0\right\rangle
 																								\left|0\right\rangle)\]
and sends all but the first system to a distant party, Bob.   The first 
system she keeps to herself.

If cloning were possible, that is if Bob had access to a machine that performs 
the operation in eq. (\ref{nonorthogonal_cloning_environ}), then application of that machine
by Bob (locally) changes the state 
\(\left|\Phi_{AB}\right\rangle \) into 
\[\left|\Phi{'}_{AB}\right\rangle = 
\frac{1}{\sqrt{2}}(\left|0\right\rangle \left|\psi_1\right\rangle \left|\psi_1\right\rangle
                                                           \left|e_{\psi_1}\right\rangle
+ \left|1\right\rangle \left|\psi_2\right\rangle \left|\psi_2\right\rangle
                                            \left|e_{\psi_2}\right\rangle).
\]
One can see that the state 
 \(\left|\Phi{'}_{AB}\right\rangle\) possesses more entanglement than 
\(\left|\Phi_{AB}\right\rangle\), for nonorthogonal 
\(\left|\psi_1\right\rangle\) and \(\left|\psi_2\right\rangle\).

\section{Discussion}

That the no-deleting and no-cloning principles follows from a principle of 
conservation of information 
 is  a remarkable result, as it allows us to understand the physical
reason for which perfect deleting or cloning are impossible. They are forbidden
because they infringe a basic law of nature - conservation of quantum
information. In particular it follows that somehow two clones means more
(subjective) information than one clone,  as opposed
to arguments in \cite{Josza_strong_cloning}. Deleting means increasing objective information
(negentropy), but
decreasing subjective information. In the case of no-cloning, we have the
converse. This is compatible with the observation that cloning increases
distinguishability, and deleting decreases it.
Let us add here that it is conceivable that suitably extended forms of the no-deleting and no-cloning principles would 
imply the information conservation principle.

It is perhaps dangerous to say that the second law is a part of an information
conservation principle. Usually the second law is associated with the arrow of time,
and with irreversibility. Conservation of information is something
completely converse.
However in our approach, there is a place for these two contradicting laws
together.
Namely, if the universe is a closed system in  a pure state, then
we can think that due to dynamics, the subsystems may get
irreversibly entangled with each other
(as discussed e.g. in \cite{HH_basic}, and consistent with the decoherence
point of view \cite{Zurek}) \cite{equilibriation,Mahler}.  
Observers have
access only to some of them, hence cannot change this entanglement.
Thus the irreversibility is here put by hands (it should follow from
features of the dynamics), and then the second law
is expressed by two facts:
\begin{enumerate}
\item[1)]  Observers can act only on part of the system
\item[2)] Information is conserved - implying, in particular,
that the local disorder (equal to entanglement \cite{entanglement}) cannot decrease.
\end{enumerate}


Similarly, when we consider efficiency of Carnot cycle, we derive
it from second law, even though the cycles are reversible.
Thus while deriving efficiences, we do not deal with irreversibility
``in action'', but rather with the \emph{implications} of irreversible
changes,
``conserved'' by information conservation: Once the entropy is increased
(which occured in the past) we cannot decrease it, and can only transfer it
somewhere else.
We could do some bad cycles, increasing entropy, but if we are interested
in optimal performance, one can keep the information  constant
(which is in practice impossible, but in principle can be done).
Going back to our picture: After  the information has been partially changed
into
entanglement with environment, one cannot regain it. The only thing one can
do is to keep
constant the information that is still accesible.


\begin{acknowledgments}

This work is supported by the University of Gda\'{n}sk, 
Grant No. BW/5400-5-0256-3 and EC grants RESQ (IST-2001-37559) and QUPRODIS (IST-2001-38877).

\end{acknowledgments}


\begin{references}
\bibitem{von_Neumann} The von Neumann entropy of a state \(\rho\), denoted as 
\(S(\rho)\), is defined to be \(-\tr\rho \log_2 \rho\).
\bibitem{unitary_deleting} This weaker form of deletion is not possible in unitary dynamics 
\cite{Braunstein_Pati_deleting}.
\bibitem{unitary_cloning}Again this weaker formulation of cloning is not possible, once we assume an 
unitary dynamics \cite{BBM}.
\bibitem{equilibriation} Let us note here that recently, in Ref. \cite{Mahler}, it was shown that 
equilibriation in thermodynamics can be explained by 
entanglement.
\end{references}
\end{document}